\documentclass[preprint,prb,amsmath,amssymb]{revtex4}
\usepackage{graphicx,color}
\usepackage{epsfig}

\newcommand{\oq}{$\omega^2$}
\newcommand{\sq}{$s^2$}
\newcommand{\wws}{$\omega_i \omega_j s_{ij}$}
\newcommand{\odo}{$\nu\omega_i \nabla^2 \omega_i$}
\newcommand{\sss}{$s_{ij}s_{jk}s_{ki}$}
\newcommand{\sds}{$\nu s_{ij} \nabla^2 s_{ij}$}
\newcommand\vis{$\nu \nabla^2 \mbox{\boldmath{$\omega$}}$}   

\begin{document}

\setlength{\baselineskip}{1.6\baselineskip} \noindent\textbf{\large
Small scale aspects of flows in proximity of the
turbulent/non-turbulent interface}

\medskip

\hspace{1cm}

\noindent M. Holzner$^1$, A. Liberzon$^2$, N. Nikitin$^3$, W.
Kinzelbach$^1$, A. Tsinober$^{2,4}$

\noindent\textit{\small $^1$ Institute of Environmental Engineering, ETH Zurich, CH 8093 Zurich, Switzerland}%
\newline
\textit{\small $^2$ Department of Fluid Mechanics and Heat Transfer,
Faculty of Engineering, Tel Aviv University, Tel Aviv 69978,
Israel}%
\newline
\textit{\small $^3$ Institute of Mechanics, Moscow
State University, 119899 Moscow, Russia}%
\newline
\textit{\small $^4$
Institute for Mathematical Sciences and Department of Aeronautics,
Imperial College, SW7 2AZ London, United Kingdom}

\medskip \hrule \medskip

{\small The work reported below is a first of its kind study of the
properties of turbulent flow without strong mean shear in a
Newtonian fluid in proximity of the turbulent/non-turbulent
interface, with emphasis on the small scale aspects. The main tools
used are a three-dimensional particle tracking system (3D-PTV)
allowing to measure and follow in a Lagrangian manner the field of
velocity derivatives and direct numerical simulations (DNS). The
comparison of flow properties in the turbulent (A), intermediate (B)
and non-turbulent (C) regions in the proximity of the interface
allows for direct observation of the key physical processes
underlying the entrainment phenomenon. The differences between small
scale strain and enstrophy are striking and point to the definite
scenario of turbulent entrainment via the viscous forces originating
in strain.} \medskip \hrule

\vspace{1cm}

Turbulent entrainment (TE) is a process of continuous transitions
from laminar to turbulent flow through the boundary (hereafter
referred to as \emph{interface}) between the two coexisting regions
of laminar and turbulent state. This process is one of the most
ubiquitous phenomena in nature and technology since, in fact, most
turbulent flows are partly turbulent: boundary layers, all free
shear turbulent flows (jets, plumes, wakes, mixing layers),
penetrative convection and mixing layers in the atmosphere and
ocean, gravity currents, avalanches and clear-air turbulence.

The first physically qualitative distinction between turbulent and
non-turbulent regions, made by Corrsin~\cite{Corrsin43,Corrsin}, is
that turbulent regions are \emph{rotational}, whereas the
non-turbulent ones are (practically) potential, thus employing one
of the main differences between turbulent flow and its random
\emph{irrotational} counterpart on the 'other' side of the
\emph{interface} separating them.

The main mechanism by which non-turbulent fluid becomes turbulent as
it `crosses' the interface is believed to involve viscous diffusion
of vorticity (\odo) at the interface~\cite{Corrsin}. Corrsin and
Kistler~\cite{Corrsin} also conjectured that the stretching of
vortex lines in the presence of a local gradient in vorticity at the
interface leads to a steepening of this gradient since the rate of
production of vorticity is proportional to the vorticity present.
The mentioned processes are associated with the small scales of the
flow. However, at large Reynolds numbers the entrainment rate and
the propagation velocity of the interface relative to the fluid are
known to be independent of viscosity (see
Ref.~[\onlinecite{Tritton,Tsinober,Hunt}] for more information and
references). Therefore the slow process of diffusion into the
ambient fluid must be accelerated by interaction with velocity
fields of eddies of all sizes, from viscous eddies to the
energy-containing eddies, so that the overall rate of entrainment is
set by the large-scale parameters of the flow. This means that
although the spreading is brought about by small eddies (viscosity),
its rate is governed by larger eddies. The total area of the
interface, over which the spreading is occurring at any instant, is
determined by these larger eddies~\cite{Tritton}.

Until recently it was difficult to implement Corrsin's distinction:
it requires information on small scale vorticity and strain which
experimentally was not accessible. This is why very little is known
about the processes at small scales and in the proximity of the
entrainment interface. A few exceptions are recent particle image
velocimetry (PIV) and planar laser-induced fluorescence (PLIF)
experiments by Westerweel~et~al.~\cite{Westerweel,WesterweelPRL} of
a jet and experiments by Holzner et al.~\cite{holzner} on
oscillating grid turbulence, in addition to the direct numerical
simulations (DNS) of a temporally developing plane wake by Bisset et
al.~\cite{Bisset} and a similar analysis of an axisymmetric
configuration by Mathew and Basu~\cite{Mathew}. Unfortunately, in
the PIV/PLIF experiments only the azimuthal component of vorticity
in a two-dimensional cross-section was accessible.


The main objective of the study presented is a systematic analysis
of the small scale dynamics associated with turbulent entrainment.
The special emphasis is on the processes involving the field of
vorticity, $\omega _{i}$, and its production/destruction by inertial
\wws\ and viscous \odo\ processes in the proximity of the interface
($s_{ij} $ are the components of the fluctuating rate-of-strain
tensor, $\nu$ is the kinematic viscosity). Studying the production
of vorticity requires access to the field of strain as well (and
thereby also to the dissipation, $2\nu s_{ij}s_{ij}$). We briefly
recall the local budget equations for enstrophy, reading
$\frac{D}{Dt}\frac{\omega^2}{2}=\omega_i \omega_j s_{ij}+\nu
\omega_i \nabla^2 \omega_i$ and strain, written as
$\frac{D}{Dt}\frac{s^2}{2}=-s_{ij}s_{jk}s_{ki}-\frac{1}{4}\omega_i
\omega_j s_{ij}-s_{ij}\frac{\partial^2 p}{\partial x_i\partial
x_j}+\nu s_{ij}\nabla^2 s_{ij}$.

Experimentally, a turbulent/non-turbulent interface was realized by
using the oscillating planar grid described in Holzner et
al.~\cite{holzner}. The grid is a fine woven screen installed near
the upper edge of a water filled glass tank, oscillating at a
frequency of 6 Hz and an amplitude of 4 mm. The scanning method of
3D particle tracking velocimetry (3D-PTV) used here is presented in
Hoyer et al.~\cite{hoyer05}. In order to access the viscous term of
vorticity, the postprocessing procedure was extended. Due to
experimental noise it is not possible to obtain the Laplacian of
vorticity, \vis, directly through differentiation as it involves
second derivatives of the vorticity field. Instead, the viscous term
is obtained from the local balance equation of vorticity in the form
$\nabla \times \mathbf{a}=\nu \nabla^2 \mbox{\boldmath{$\omega$}}$
by evaluating the term $\nabla \times \mathbf{a}$ from the
Lagrangian tracking data. The derivatives of Lagrangian
acceleration, $\partial a_{i} / \partial x_{j}$, are calculated in
the same way as the derivatives of the velocity, $\partial u_{i} /
\partial x_{j}$, described in Hoyer et al.~\cite{hoyer05} and
subsequently they are interpolated on a Eulerian grid. The number of
tracked particles is about 6$\cdot$10$^3$ in a volume of
2$\times$2$\times$1.5~cm$^{3}$ and the interparticle distance is
about 1~mm, which is slightly above the estimated Kolmogorov length
scale, $\eta$=0.6mm. The Taylor microscale, $\lambda$, is about
7~mm. The spacing of the Eulerian grid was taken equal to the
interparticle distance. Further details on the data analysis will be
discussed in the forthcoming full paper. In both experiment and
simulation, the Taylor microscale Reynolds number is
Re$_{\lambda}$=50.

Direct numerical simulation (DNS) was performed in a box
(side-lengths $L_1$, $L_2$, $L_3$) of initially still fluid. Random
(in space and time) velocity perturbations are applied at the
boundary $x_2$=0. The procedure of generating the boundary
conditions is as follows. For a fixed time and in the discrete set
of points, $x_1 = k\Delta_l$, $x_3 = l\Delta_l$ (k, l
- integers), each velocity component, $u_i$ (i = 1, 2, 3), is calculated as $%
u_i = V_i\xi$, where $\xi$ is a random number within the interval
$[-1,1]$ and $V_i$ is a given velocity amplitude. For other times
and spatial points ($x_1$, $x_3$) boundary velocities are obtained
by cubic interpolation in time and bilinear interpolation in space.
At each time the three boundary velocity components yield zero
average value over the boundary plane. The method of boundary
velocity assignment determines the velocity scale, $V =
\mathrm{max}(V_i)$ and the length scale $\Delta_l$. Together with
the viscosity of a fluid, $\nu$, these parameters define the
Reynolds number $Re = V \Delta_l/\nu=1000$ of the simulation. The
Navier-Stokes equations are solved with periodic boundary conditions
for the directions $x_1$ and $x_3$, with periods $L_1$ and $L_3$,
respectively. The computational domain is finite in the $x_2$
direction, as $x_2 \leq L_2$. Shear-free conditions $\partial
u_1/\partial x_2=\partial u_3/\partial x_2=u_2=0$ are imposed at the
boundary $x_2$ = $L_2$. A mixed spectral-finite-difference method is
used for the spatial discretization and the time advancement is
computed by a semi-implicit Runge-Kutta method~\cite{Nikitin94}. The
resolution is 192$\times$192 Fourier modes in $x_1$ and $x_3$
directions and 192 grid points in $x_2$ direction. The local
Kolmogorov length scale is twice the grid spacing.


\begin{figure}[ht]
\centering
\includegraphics[width=\columnwidth]{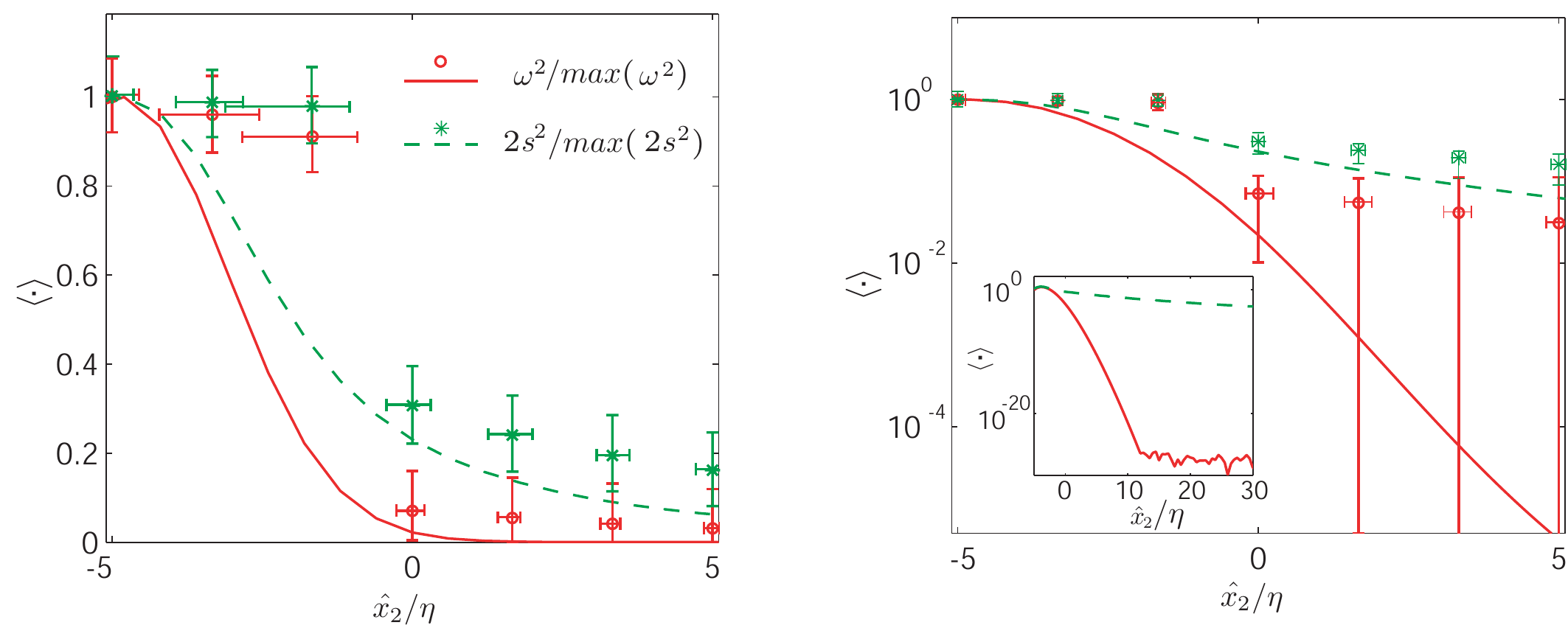}
\caption{Average profiles of $\protect\omega^2(\hat{x}_2)$ and
$2s^2(\hat{x}_2)$ from PTV (symbols), and DNS (lines) relative to
the interface, $\hat{x}_2 = x_2 - x_2^*$, on linear scale (left) and
on log-scale (right). The values are normalized by the respective
maxima ($\mathrm{max}$(\oq)$\approx\mathrm{max}$(2s$^2)\approx$3.6
s$^{-2}$ for PTV and 3.0 for DNS). The horizontal error bars display
the sensitivity of the experimental result on the enstrophy
threshold, 0.5-2.5 $s^{-2}$, the vertical error bars represent the
accuracy of the measurement. The symbol $\langle \cdot \rangle$
denotes the ensemble average of the respective quantity}
\label{fig:cond_mean_ptv_dns}
\end{figure}

In both experiment and simulation, turbulence is generated at the
plane $x_2$ =0 and propagates along $x_2>0$. Firstly, the interface
is identified at $ x_2^*(t)$, using a fixed threshold of enstrophy,
(for details see Ref.~[\onlinecite{holzner}] and references therein)
and the analysis is done with respect to the interface location, as
in Fig.~1, in which profiles of enstrophy,
$\omega^2=\omega_i\omega_i$ and strain rate, $s^2=s_{ij}s_{ij}$,
averaged over homogeneous $x_1, x_3$ directions on linear scale
(left) and log scale (right), are shown. The distance to the
interface, $\hat{x}_2$ is normalized by the Kolmogorov length scale,
$\eta$. The 'proximity' or 'region of the interface' hereafter
refers to the interval -5$<\hat{x}_2/\eta<$5. We observe that the
rate of strain on the non-turbulent side of the interface remains
high in contrast to enstrophy which drops much more steeply.
Experimentally it is not possible to obtain enstrophy lower than a
small (but finite) level of noise. This is one of the reasons why
comparison to DNS is presented for all the results. In the DNS, the
numerical noise level is reached at $\hat{x_2}/\eta>$10 and this
level is about 25 decades lower in magnitude, see the inset in
Fig.~1 (right).

\begin{figure}[ht]
\centering
\includegraphics[width=\columnwidth]{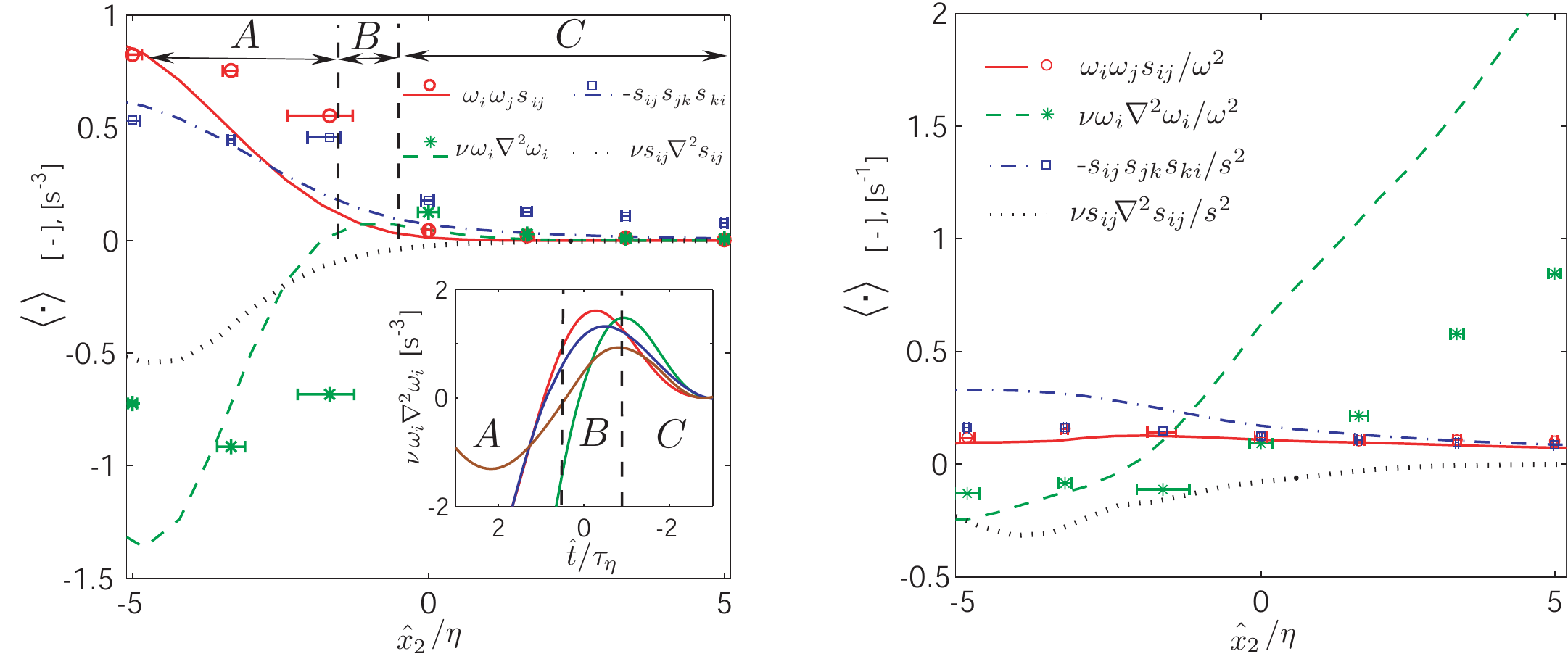}
\caption{Average profiles of production and viscous destruction
terms of strain and enstrophy (left) and their rates (right). The
inset shows the
individual Lagrangian trajectories of $\protect\nu\protect\omega_i \protect%
\nabla^2 \protect\omega_i\ $ obtained from PTV. Lines are from DNS, symbols
are from PTV}
\label{fig:cond_mean_enstrophy_rates}
\end{figure}

Fig.~2 shows profiles of 
production and viscous terms of strain and enstrophy (left) and
their rates (right). We note that the viscous term, \odo, exhibits a
remarkable behavior showing a distinct maximum in the region of the
interface. In addition, the individual Lagrangian trajectories
(examples are shown in the inset in
Fig.~2, where the abscissa is 
$\hat{t}=t-t^{\ast }$, normalized by the Kolmogorov time scale,
$\tau_{\eta } $, and $t^{\ast }$ is analogous to $x_{2}^{\ast }$)
possess such an extremum. Therefore, we use the maximum of the
viscous term as the exact location of the interface, defined in a
physically more appealing way than the threshold-dependent crossing
of $\hat{x}_{2}/\eta $=0. For the further analysis we define three
physically distinct regions of the interface with respect to the
maximum of \odo\ (marked in
Fig.~2): (A) the \emph{turbulent} 
region, in which the behavior of the viscous term is `normal', i.e.
it is negative in the mean, (B) the interval between the peak and
the point where $\langle $\odo$\rangle $=0 is termed
\emph{intermediate} region (with the 'abnormal' viscous production)
and, (C) the \emph{non-turbulent} region from the peak to
$\hat{x}_{2}/\eta $=5. The positiveness of both \wws\ and \odo\ is a
peculiar feature of the regions B and C, in contrast to region A,
where, in the mean, \odo\ contributes to the destruction and \wws\
to the production of \oq. It is noteworthy that strain behaves
rather differently from vorticity. In particular, the viscous term,
\sds, is negative in the mean in all three regions, i.e. it is not
building up \sq. In Fig.~2 we see 
also that strain production, -\sss, is significant and it is (in the
mean) not balanced by \sds\ in region C. When the rates of
quantities are considered it appears
(Fig.~2, right) that the role of 
viscous production is even more important: the term \odo/\oq\
attains high positive values in region~C, decreases along region~B
and finally becomes negative in region~A (balancing the average
\wws/\oq). In contrast, the term \wws/\oq\ and the analogous rates
of the strain viscous and production terms do not change as
drastically and remain of the same sign.

\begin{figure}[ht]
\centering
\includegraphics[width=\columnwidth]{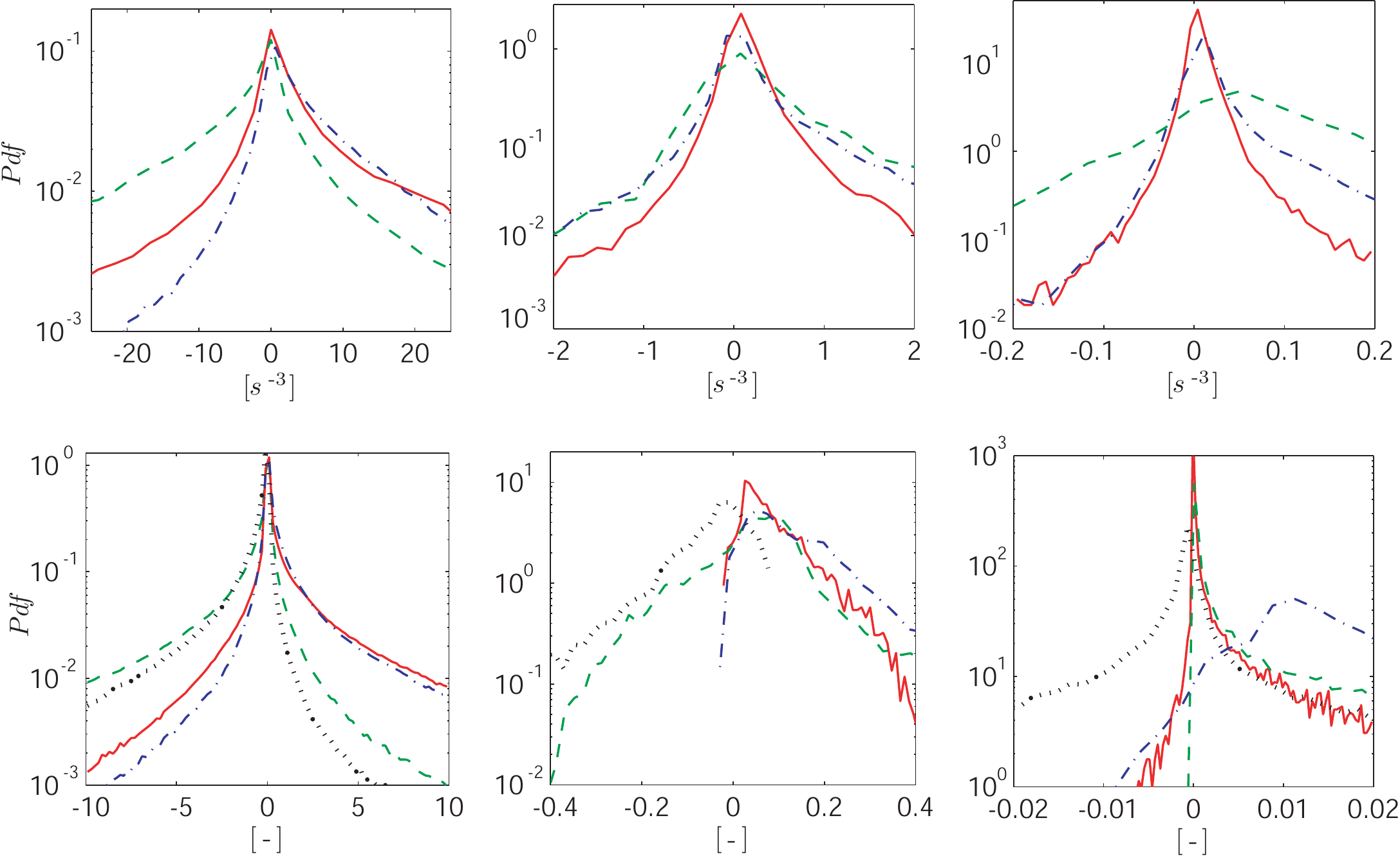}
\caption{PDF of various quantities from experiment (top) and
simulation (bottom) according to the division in 3~regions:
turbulent (left), intermediate (center) and non-turbulent (right).
$\protect\omega_i \protect\omega_j s_{ij}\ $
(--), \odo\ (- -), -$%
s_{ij}s_{jk}s_{ki}\ $ (- $\cdot$), \sds\ ($\cdots$,~only bottom)}
\label{fig:pdf_enstrophy_balance}
\end{figure}

Fig.~3 presents the estimations of 
probability density functions (PDFs) of the relevant terms from the
different regions (A,B, and C, from left to right; PTV top, DNS
bottom). Consistently with the other results, the PDFs of both \wws\
and \odo\ are positively skewed in regions B and C. In region B we
note that the probability of negative events of \odo\ and positive
events of \wws\ increases as compared to region C. Finally, as
expected, in region A the PDF of \odo\ is negatively skewed. The
changes of the strain production and viscous terms between the
regions A-C are less drastic. Essentially, -\sss\ is positively and
\sds\ is negatively skewed in all the three regions.

\begin{figure}[ht]
\centering
\includegraphics[width=\columnwidth]{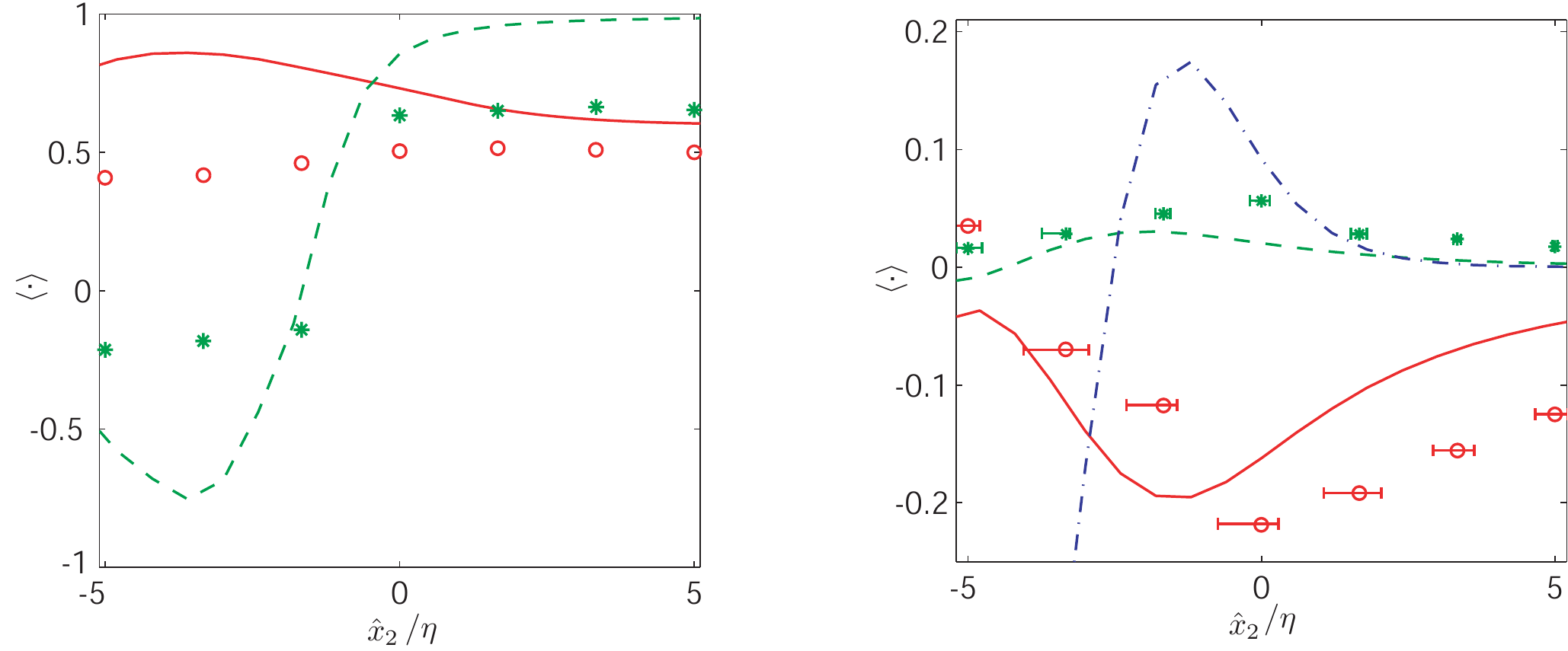}
\caption{(left) Cosines of the angle between vorticity and
\emph{\textbf{W}} (--,o) and between vorticity and \vis\ (-
-,$\ast$). (right) Invariants, $Q$ (--, o), $R$ (- -,$\ast$), $S$
($- \cdot -$,~only DNS). Lines are from DNS, symbols are from PTV}
\label{fig:cosines_vorticity}
\end{figure}

For the understanding of the interaction of strain and enstrophy in
the proximity of the interface it is very instructive to look at the
invariants of the gradient tensor: $Q=\omega ^{2}-2s^{2}=-\nabla
^{2}P$ shows the relative strength of vorticity and strain,
$R$=-1/3($s_{ij}s_{jk}s_{ki}\ $+3/4$\omega _{i}\omega _{j}s_{ij}\ $)
relates to their production terms, and $S$ is the quantity related
to the two viscous terms, $S=\nu \omega _{i}\nabla^{2}\omega _{i}\
-2\nu s_{ij}\nabla ^{2}s_{ij}\ $ (Fig.~4, 
right). Since the mean values of $Q,R$ and $S$ vanish identically
for homogeneous turbulence, their nonzero values indicate the degree
of inhomogeneity in the proximity of the interface. Apparently,
inhomogeneity is the property which is maximal where also \odo\ is
maximal. In the same context it is also interesting to look at the
cosine of the angle between vorticity and its Laplacian, $\nabla
^{2}\omega _{i}$, shown in Fig.~4 (left), 
which exhibits significant changes across the regions A,B, and C.
The observed transition from positive (alignment) to negative
(anti-alignment) values is in agreement with the qualitatively
different behavior of \odo\ in these regions. In contrast to that,
the alignment between vorticity and the vortex stretching vector,
$W_i=\omega_{j}s_{ij}$, changes only weakly throughout regions A-C
and is consistent with the positiveness of $\langle $\wws$\rangle$
mentioned above. The results indicate that an interpretation of the
viscous term \odo\ as \emph{interaction between strain and vorticity
due to viscosity} (i.e. due to the curl of the viscous force
originating from the divergence of the strain tensor) is physically
more appealing than 'simple' diffusion of vorticity due to
viscosity. We emphasize that \odo\ is the interaction of vorticity
and strain since (e.g., Ref.~[\onlinecite{Batchelor}]) \vis$=1/\rho\
\nabla\times \mathbf{F}^{s}$, where $\ F_{i}^{s}=2\nu\partial
/\partial x_{k}\{s_{ik}\}$ and $\rho$ is the fluid density.

In summary, we analyzed small scale enstrophy and strain dynamics in
proximity of a turbulent/non-turbulent interface without strong mean
shear. The experimental results are in good agreement with the
simulation, at least on a qualitative level, which is considered as
a clear indication for the reliability of both methods. The behavior
of vorticity-related quantities is very different from the
strain-related counterparts. For example, the viscous term is not
responsible for building up strain as strain is destroyed by \sds\
in all three regions. In addition, the analysis of these quantities
with respect to the distance from the interface reveals the range of
influence of $\nu \omega _{i}\nabla ^{2}\omega _{i}\ $ and $\nu
s_{ij}\nabla ^{2}s_{ij}\ $ into the non-turbulent region. We also
found that both \wws\ and \odo\ are responsible for the increase of
\oq\ at the interface and substantiate the physical interpretation
of the term \odo\ as \emph{viscous interaction}, in analogy to \wws,
commonly referred to as the \emph{inviscid interaction} of vorticity
and strain.

\medskip

We gratefully acknowledge the support of this work by ETH Grant No.
0-20151-03. The work of N. Nikitin was supported by the Russian
Foundation for Basic Research under the grant 05-01-00607.

\end{document}